\documentclass[aps,prx,reprint,superscriptaddress,footnote, twocolumn,notitlepage]{revtex4-1}

\usepackage{amsmath, amsthm, amssymb,amsfonts,mathbbol,amstext,xparse}
\usepackage{pifont}
\usepackage{graphicx}
\usepackage{dcolumn}
\usepackage{bm}
\usepackage{bbm}
\usepackage{hyperref}
\usepackage{mathtools}
\usepackage{comment}
\usepackage{color}
\usepackage{multirow}
\usepackage{diagbox}
\usepackage{float}
\usepackage{xcolor}
\usepackage{makecell}
\usepackage[normalem]{ulem}

\def\1{\mathbf{1}}
\def\0{\mathbf{0}}

\DeclareMathOperator{\tr}{tr}

\newcommand{\ket}[1]{| #1 \rangle}
\newcommand{\bra}[1]{\langle #1 |}
\newcommand{\braket}[2]{\left\langle #1 \mid #2 \right\rangle}

\newcommand{\processnext}[1]{%
\ifx\listfinish#1\empty\else\listact{#1}\expandafter\processnext\fi}

\newcommand{\eqnref}[1]{(\ref{#1})}

\begin{document}

\title{Measurement incompatibility under loss}
	\author{Mohammad Mehboudi}
	\email{mohammadmehboudi@gmail.com}
        \address{Technische Universität Wien, 1020 Vienna, Austria}
        \author{Fatemeh Rezaeinia}\email{f.rezayinia@gmail.com}
        \address{Department of Physics, University of Tehran, P.O. Box 14395-547, Tehran, Iran}
	\author{Saleh Rahimi-Keshari}
	\email{s.rahimik@gmail.com}
	\address{School of Physics, Institute for Research in Fundamental Sciences (IPM), P.O. Box 19395-5531, Tehran, Iran}
 \address{Department of Physics, University of Tehran, P.O. Box 14395-547, Tehran, Iran}
\date{\today}
\begin{abstract}
We investigate the measurement incompatibility of continuous-variable systems with infinite-dimensional Hilbert spaces under the influence of pure losses, a fundamental noise source in quantum optics, and a significant challenge for long-distance quantum communication. We show that loss channels with transmissivities less than $1/n$ make any set of $n$ measurements compatible. 
Additionally, we design a set of measurements that remains incompatible even under extreme losses, where the number of measurements in the set increases with the amount of loss. 
These measurements rely on on-off photodetectors and linear optics, making them feasible for implementation under realistic laboratory conditions. Furthermore, we demonstrate that no loss channel can break the incompatibility of all measurements. As a result, quantum steering remains achievable in the presence of pure loss.

\end{abstract}

\maketitle

{\it Introduction.---}Measurement incompatibility remains a linchpin of quantum information processing, as it is necessary to violate any steering or Bell inequalities, and is required to perform several quantum information processing tasks such as quantum key distribution or communication tasks~\cite{RevModPhys.95.011003,RevModPhys.86.419,RevModPhys.92.015001,RevModPhys.92.025002,Gisin2007}. In the early stages of the development of quantum mechanics it was understood that it is impossible to simultaneously measure certain properties of a quantum system, like position and momentum, with arbitrary precision \cite{Heisenberg1925,Robertson1929}. Measurement incompatibility generalizes the idea of non-commutativity of observables---or projective rank-one measurements---to 
generalized measurements described by positive operator valued measures (POVM). Consider a quantum system living in a Hilbert space ${\cal H}$. Also, consider a set of $n$ measurements $\{\{M_{\bm a_j}^j\}_{\bm a_j}\}_{j=1}^{n}$ that live within this Hilbert space, where $M_{\bm a_j}^j$ is a POVM element of the $j$th measurement representing the outcome $\bm a_j$, which can be continuous or discrete. The POVM elements should satisfy $M_{\bm a_j}^j\geq 0~\forall \{j,\bm a_j\}$, and $\sum_{\bm a_j}M_{\bm a_j}^j=I ~~\forall j$, with $I$ being the identity operator in the same Hilbert space. For continuous measurements, one should instead replace $\sum \to \int d\bm a_j$.
The set of measurements is said to be compatible if and only if it admits the following decomposition
\begin{align}\label{eq:compatible_def}
    M_{\bm a_j}^j = \sum_{\bm a_{k\neq j}} M_{\bm a_1,\dots,\bm a_n},~~\forall \{j,\bm a_j\},
\end{align}
such that $\{M_{\bm a_1,\dots,\bm a_n}\}_{\bm a_1,\dots,\bm a_n}$ forms a valid POVM, which is termed the parent (or mother) measurement for the set. Hereafter, we use  $\vec{\bm a}$ for ${\bm a_1,\dots,\bm a_n}$ to shorten our notation. Any set of measurements that does not admit a form like Eq.~\eqref{eq:compatible_def} is said to be incompatible.

Besides its fundamental importance, measurement incompatibility is also important from a practical point of view. The concept of measurement incompatibility is central to the study of quantum entanglement and its applications in quantum teleportation and quantum computing \cite{Horodecki2009}. This property is essential for many quantum information tasks like quantum steering and violation of Bell inequalities, as the statistics of a set of compatible measurements can be fully described by local hidden state models~\cite{Quintino2014,Uola2014}. In a quantum state discrimination task, the utilization of sets of incompatible measurements confers a distinct advantage relative to employing compatible measurements \cite{Skrzypczyk2019QSD,Carmeli2019QSD,Sen2024QSD}. Thus, it plays a crucial role in quantum communication \cite{Saha2023communication}. It is exploited to achieve secure communication protocols based on the principles of quantum key distribution~\cite{Gisin2002}. Additionally, in quantum state tomography, incompatible measurements are used to reconstruct the complete quantum state of a system~\cite{paris2004quantum,PRXQuantum2022}. Thus, a deep understanding of measurement incompatibility is essential for leveraging the full potential of quantum systems in information processing tasks.
Given its practical relevance, measurement incompatibility has attracted significant attention in recent years. Two key questions arise in this area: first, how to determine when a set of measurements is incompatible, and second, how errors affect incompatibility and subsequently to assess their robustness against imperfections.

Numerous studies focus on finite-dimensional systems when investigating measurement incompatibility~\cite{Skrzypczyk2015oneNth,Heinosaari2015Noise,Uola2016JM,Bavaresco2017,Designolle2019Quantifying,Uola2021subspace,Jones2023HighD}, as these systems are easier to handle both theoretically and computationally. For instance, semi-definite programming (SDP) can efficiently determine the compatibility of finite-dimensional measurements and identify parent measurements~\cite{skrzypczyk2023SDP}. However, many settings, particularly in quantum optics, engage with infinite-dimensional systems. This makes the study of infinite-dimensional systems crucial. 
Furthermore, analyzing the impact of imperfections on measurement incompatibility requires studying how quantum channels affect measurements. Most previous work has focused on the Gaussian regime, examining Gaussian states and measurements, or those with specific quasi-probability distributions. For instance, it is known that under $50\%$ losses incompatibility of all measurements with non-negative Wigner function (including all Gaussian measurements) breaks~\cite{Heinosaari2015Gaussian,PhysRevA.104.042212}. However, an important question concerns the effects of errors on general non-Gaussian measurements. In this regard, Ref.~\cite{Ji2016steering}, demonstrates the violation of a steering inequality using three non-Gaussian measurements even under losses above $50\%$. Given the link between steering and measurement incompatibility~\cite{PhysRevA.93.052115,Uola2014,PhysRevA.93.052112}, this suggests that these non-Gaussian measurements remain incompatible beyond $50\%$ loss.

In this paper, we study the effect of loss on measurement incompatibility. We demonstrate that any set of $n$ arbitrary measurements becomes compatible under a lossy channel with transmissivity $ \tau \leq 1/n $. However, increasing the number of measurements to $n+1$ can maintain incompatibility under $\tau \geq 1/n$. Specifically, a set of displaced on-off photodetection measurements provides an example of $n+1$ measurements that remain incompatible under $ \tau \geq 1/n $. We prove this claim numerically for $2\leq n \leq 14$ and conjecture it to hold for arbitrary $n$. These measurements can be implemented in the laboratory with technologies in hand. Furthermore, using these same measurements, we show that there is no loss channel that breaks the incompatibility of all measurements; also implying that there exists no loss channel that breaks the quantum steerability of all states. Therefore, our results provide a practical solution to overcome loss, which is a common error in experimental implementations of quantum information and quantum computing tasks.

{\it The loss channel.---}The pure loss can be completely characterized by its operation on coherent states
\begin{align}\label{eq:loss-def}
    {\cal E}_{\tau}(\ket{\alpha}\bra{\alpha}) \coloneqq \ket{\sqrt{\tau}{\alpha}}\bra{\sqrt{\tau}{\alpha}},
\end{align}
where $0\leq\tau\leq 1$ is the transmissivity parameter, with $\tau=1$ representing a lossless channel, while $\tau=0$ means vacuum output state for any input. The simplest model for a loss channel is a beam splitter with transmissivity $\tau$ whose second input port is always in the vacuum state and the second output mode is lost to the environment. Specifically, we can describe the action of a loss channel on state $\rho$ as $\smash{{\cal E}_{\tau}(\rho)=\tr_2 \big(\mathcal{U}_{\rm BS} \rho \otimes \ket{0}\bra{0} \mathcal{U}
^{\dagger}_{\rm BS}\big)}$. The unitary operator $\mathcal{U}_{\rm BS}$ describes the beam splitter, and its action on coherent and vacuum states $\mathcal{U}_{\rm BS}\ket{\alpha}\ket{0}= \ket{\sqrt{\tau} \alpha}\ket{\sqrt{1-\tau} \alpha}$ gives the definition \eqnref{eq:loss-def}. 

The $m$-mode generalization of the beam splitter transformation is a linear-optical network (LON) that transforms coherent states to coherent states, $\mathcal{U}_{\rm LON} \ket{\alpha_1,\dots,\alpha_m}=\ket{\beta_1,\dots,\beta_m}$, where $\beta_k=\sum_{j=1}^{m} U_{jk} \alpha_j$ with $U$ being the unitary transfer matrix associated with the LON. Let us define the LON channel ${\mathcal E}_{\rm LON}^{(m)} : {\mathcal H} \mapsto {\mathcal H}^{\otimes m}$ as ${\mathcal E}_{\rm LON}^{(m)}(\rho)=\mathcal{U}_{\rm LON} \rho\otimes  \ket{0^{m-1}}\bra{0^{m-1}} \mathcal{U}_{\rm LON}^\dagger$,  which means that an input state $\rho$ is injected into the first input mode of an $m$-mode LON and all other input modes are in their vacuum states. In this picture, the loss channel~\eqref{eq:loss-def} can generally be thought of as a single-mode marginal channel of ${\mathcal E}_{\rm LON}^{(m)}$ defined as ${\mathcal E}_{\tau_k}(\rho)=\tr_{k^c}\!\big({\mathcal E}_{\rm LON}^{(m)}(\rho)\big)$, where the corresponding LON matrix element describing the connection between the first input and the $k$th output modes is $U_{1k}=\sqrt{\tau_k}$, and the trace is over all output modes except $k$. 


{\it Result 1.---} The set of $n$ loss channels $\{\mathcal{E}_{\tau_k} \}_{k=1}^n$ can be viewed as the single-mode marginal channels of an LON channel ${\mathcal E}_{\rm LON}^{(m)}$ with $n\leq m$, if ${\sum_{i=1}^n \tau_i} \leq 1$. We also show that there cannot exist a quantum channel ${\cal E} : {\cal H} \mapsto {\cal H}^{\otimes n}$ with marginal channels $\{\mathcal{E}_{\tau_k} \}_{k=1}^n$ if ${\sum_{i=1}^n \tau_i} > 1$.

{\it Proof of Result 1.---}For the first part, it is easy to check that if ${\sum_{i=1}^n \tau_i} \leq 1$ we can always find an LON with transfer matrix elements $U_{1k}=\sqrt{\tau_k}$, as any arbitrary normalized vector can be considered as the first row of an LON transfer matrix. Therefore, we have $\mathcal{E}_{\tau_k}(\ket{\alpha}\bra{\alpha})=\tr_{k^c}\!\big(\mathcal{U}_{\rm LON} \ket{\alpha}\bra{\alpha} \otimes  \ket{0^{m-1}}\bra{0^{m-1}} \mathcal{U}_{\rm LON}^\dagger\big)=\ket{\sqrt{\tau_k}{\alpha}}\bra{\sqrt{\tau_k}{\alpha}}$. Note that $m=n$ for ${\sum_{i=1}^n \tau_i} = 1$. A proof of the second part is by contradiction. Suppose that such a channel $\mathcal{E}$ exists, then for an input coherent state $\ket{\alpha}$ it generates the tensor product of coherent states $\otimes_{k=1}^{n}\ket{\sqrt{\tau_k} {\alpha}}$. Then by applying a beam splitter transformation on each coherent state and an ancillary vacuum state, we have $\otimes_{k=1}^{n} \mathcal{U}_{\rm BS,k}(\ket{\sqrt{\tau_k} {\alpha}}\ket{0})=\otimes_{k=1}^{n}\ket{\sqrt{\eta_k\tau_k} {\alpha}} \otimes_{k=1}^{n}\ket{\sqrt{(1-\eta_k)\tau_k} {\alpha}}$, where the beam splitter transmissivities $\eta_k$ are chosen such that $\sum \eta_k \tau_k=1$. Now consider an LON with the transfer matrix whose first column is $U_{j1}=\sqrt{\eta_j\tau_j}$. We can simply see that acting this LON on the $n$-mode coherent state $\otimes_{k=1}^{n}\ket{\sqrt{\eta_k\tau_k} {\alpha}}$ gives $\ket{\alpha}\ket{0^{n-1}}$. Putting everything together, using the channel $\mathcal{E}$ and further beam splitter and LON operations, we could transform any \textit{arbitrary} single-mode coherent state $\ket{\alpha}$ to the $(1+n)$-mode state $\ket{\alpha} \otimes_{k=1}^{n}\ket{\sqrt{(1-\eta_k)\tau_k} {\alpha}}$. However, this transformation is not allowed by quantum mechanics. Repeating this transformation and combining states $\otimes_{k=1}^{n}\ket{\sqrt{(1-\eta_k)\tau_k} {\alpha}}$ using an LON enable us to prepare copies of coherent states that are not orthogonal to one another; this violates the no-cloning theorem. Therefore, loss channels $\{\mathcal{E}_{\tau_k} \}_{k=1}^n$ with ${\sum_{i=1}^n \tau_i} > 1$ cannot be single-mode marginal channels of any quantum channel.

\begin{figure}
  \includegraphics[width=0.8\columnwidth]{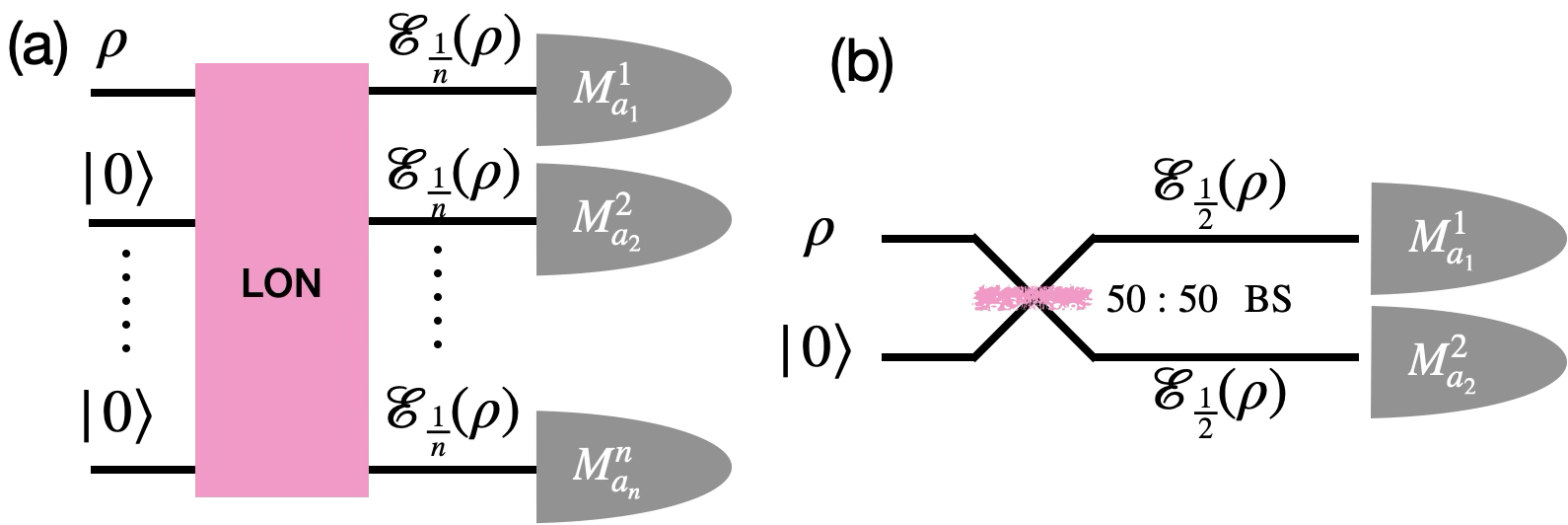}
  \caption{The lossy channel can be always represented as marginals of a passive LON. (a) If $\tau=1/n$, an LON exists that takes the signal and $n-1$ inputs in vacuum, and its $n$ marginal outputs are all ${\cal E}_{\frac{1}{n}}(\rho)$. Such amount of loss breaks incompatibility of \textit{any} set of $n$ measurements. (b) The case of $\tau=1/2$ as an example can be simulated by an additional input in the vacuum state that interacts with the signal on a balanced beam splitter. Both outputs are locally equal. This breaks the incompatibility of any pair of measurements.}
  \label{fig:LOSS_BS}
\end{figure}


{\it Result 2.---}A loss channel with transmissivity $\tau \leq 1/n$ breaks the incompatibility of any arbitrary set of $n$ measurements $\{\{M_{\bm a_j}^j\}_{\bm a_j}\}_{j=1}^{n}$ . Specifically, denoting ${\cal E}_{\tau}^*$ as the dual of the loss channel defined through $\tr \big(M_{\bm a_j}^j {\mathcal E}_{\tau}(\rho)\big) \eqqcolon \tr\big(\rho\, {\mathcal E}_{\tau}^*(M_{\bm a_j}^j)\big)$, the set of measurements $\{\{{\mathcal E}_{\tau}^*(M_{\bm a_j}^j)\}_{\bm a_j}\}_{j=1}^{n}$ with $n\leq \lfloor{1/\tau}\rfloor$ is jointly measurable---with $\lfloor{\circ}\rfloor$ being the floor function. 

{\it Proof of result 2.---} Let us first consider the case of $\tau=1/n$. Using \textit{Result 1}, we can see that $n$ loss channels with transmissivities $\tau=1/n$ can be viewed as the marginal channels of an LON channel. As shown in Fig.~\ref{fig:LOSS_BS}, using a passive LON with a transfer matrix whose first row is given by $U_{1k}=\sqrt{\tau}$ for $k=1,\dots,n$, followed by a measurement at each output mode, we can construct the parent measurement for any $n$ measurements after a loss channel with $\tau=1/n$. Therefore,  the set of measurements $\{\{{\mathcal E}_{1/n}^*(M_{\bm a_j}^j)\}_{\bm a_j}\}_{j=1}^{n}$ become compatible and the POVM elements of the parent measurement are given by 
\begin{equation}\label{eq:parent_LON}
      {\tilde M}_{\vec {\bm a}} \coloneqq \bra{0^{n-1}}\, \mathcal{U}_{\rm LON}^\dagger M_{{\bm a}_1}^1\otimes \cdots \otimes M_{{\bm a}_n}^n  \mathcal{U}_{\rm LON}\ket{0^{n-1}}.
\end{equation}
To verify that indeed $\sum_{\bm a_{k\neq j}} {\tilde M}_{\vec {\bm a}}=\mathcal{E}^*_{1/n}(M_{{\bm a}_j}^j)$, one can simply show that for arbitrary $\rho$ we have $\tr \big(\rho \sum_{\bm a_{k\neq j}}  {\tilde M}_{\vec {\bm a}}\big)=\tr\big(\mathcal{E}_{1/n}(\rho)M_{{\bm a}_j}^j\big)$ using the above expression (see Appendix\ref{app:marginals}). As an example, we can see that a loss channel with $\tau=1/2$ makes any two arbitrary measurements compatible. In this case, the parent measurement is constructed using a balanced beam splitter as shown in Fig.~\ref{fig:LOSS_BS}.

For the case of $\tau<1/n$, we can write $\tau=\eta/n$ with $0<\eta<1$. We can still see that any $n$ measurements become compatible. In this case, the parent measurement can be constructed by adding a loss channel with transmissivity $\eta$ before the parent measurement in the above case. Therefore, the POVM elements of the parent measurement are $\mathcal{E}^*_\eta(M_{\vec {\bm a}}^n)$, which can be verified by
\begin{align*}
    \sum_{\bm a_{k\neq j}}\! \mathcal{E}^*_\eta({\tilde M}_{\vec {\bm a}}) =\mathcal{E}^*_\eta\Big(\!\sum_{\bm a_{k\neq j}}{\tilde M}_{\vec {\bm a}}\Big)\!=\mathcal{E}^*_\eta\big(\mathcal{E}^*_{\frac 1 n}(M_{{\bm a}_j}^j)\big)=\mathcal{E}^*_{\frac \eta n}(M_{{\bm a}_j}^j).
\end{align*}
Here we used the fact that two successive loss channels with transmissivities $\tau_1$ and $\tau_2$ can be viewed as a loss channel with transmissivity $\tau_1 \tau_2$. This property of loss channels can be verified using the definition~\eqref{eq:loss-def}.

Having shown the effect of loss in making any $n$ measurements compatible, we now ask if there exists a set of $n+1$ measurements that remain incompatible under $\tau=1/n$. In the following, we prove this to be the case for 
$n\in\{1,\dots,14\}$ by construction and conjecture it to be true for arbitrary $n$.

{\it Result 3.---}There exist a set of $n+1$ measurements that remain incompatible under $\tau = 1/n$ transmissivity. We prove this result numerically for $2\leq n \leq 13$, and conjuncture it to hold for arbitrary $n$.

\begin{figure}
    \centering
    \includegraphics[width=.8\columnwidth]{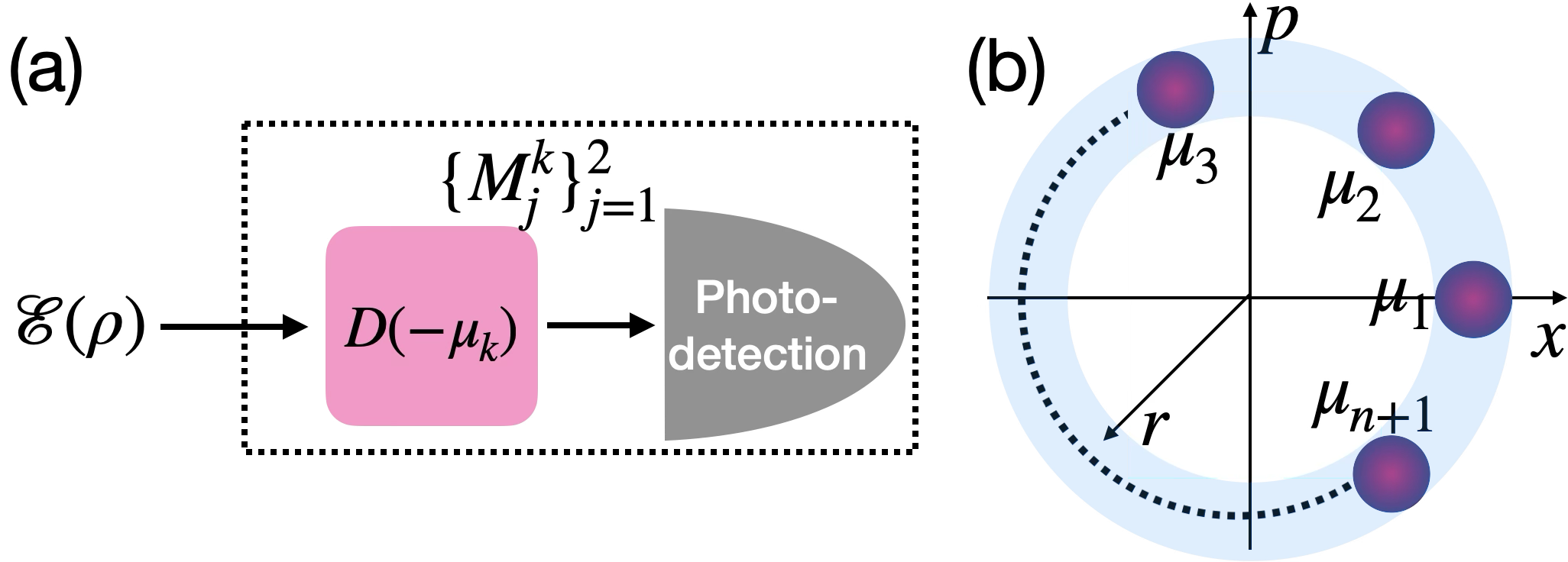}
    \caption{(a) Our proposed measurements are similar to on-off photodetection, with the difference that instead of projection into the vacuum state projects into different coherent states. This can be realized by the action of a displacement unitary followed by a standard photodetection. (b) These coherent states are chosen such that they have the same distance $r$ from the origin of the phase-space, and are separated from one another with a $2\pi/(n+1)$ phase.}
    \label{fig:displaced_pd}
\end{figure}

{\it Proof of Result 3.---}Our proof is by construction. The set of measurements that we propose for this task are implementable by simple means, namely 
displacement operation and on-off photodetection. To begin with, take the following two-outcome measurement with POVM elements 
\begin{align}\label{eq:disp_photo_det}
  M_{1}^k = \ket{\mu_k}\bra{\mu_k},
  \hspace{1cm}M_{2}^k = I-M_{1}^k,
\end{align}
with $\ket{\mu_k}$ being a coherent state. The panel (a) in Fig.~\ref{fig:displaced_pd} shows how this can be implemented in practice by firstly applying the displacement operation  $D(-\mu_k)$ on the quantum state followed by a standard on-off photodetection. The displacement operation can be realized by overlapping the state and a coherent state on a highly transmissive beam splitter~\cite{Kuhn2018}. One can show that for any two $\mu_1\neq \mu_2$, the set $\left\{\{M_j^1\}_j,\{M_k^2\}_k\right\}$ is incompatible. We leave the proof to the Appendix~\ref{app:2_disp}. 
Furthermore, in Appendix~\ref{app:2_disp_tau} we also show that by choosing $\mu_1 = r = -\mu_2$ and taking the limit of $r\to 0$, the two measurements remain incompatible for any $\tau > 1/2$. That is, the two measurements will lose their incompatibility only under 
$\tau \leq 1/2$, which also breaks the incompatibility of any other pair of measurements. This suggests that these measurements are very robust against loss, despite their simplicity. 

Inspired by these properties, we now consider the set of $n+1$ measurements $\{\{M_{j}^k\}_{j=1}^2\}_{k}$ with outcomes given in  Eq.~\eqref{eq:disp_photo_det}.

We choose $\mu_k = r\exp\left[i{\frac{2(k-1) \pi}{n+1}}\right]$, with $k\in\{1,\dots,n+1\}$---see also panel (b) of Fig.~\ref{fig:displaced_pd} for the phase-space representation.
To proceed further, we need to see how the measurements are affected by loss. Computing the $Q$ function~\cite{Husimi,Cahill-Glauber1969} of operator $\mathcal{E}^*_\tau(M_{1}^k)$,
\begin{equation}\label{eq:Q-function}
    Q\big(\alpha|\mathcal{E}^*_\tau(M_{1}^k)\big)=\frac{1}{\pi} \bra{\alpha}\mathcal{E}^*_\tau(M_{1}^k)\ket{\alpha}=\frac{1}{\pi} e^{-\tau\big|\alpha-\frac{\mu_k}{\sqrt{\tau}}\big|^2}\! ,
\end{equation}
where we used the definition \eqref{eq:loss-def} and $|\bra{\sqrt{\tau}\alpha}\mu_k\rangle|^2=\exp(-|\sqrt{\tau}\alpha-\mu_k|^2)$, shows that $\tau \mathcal{E}^*_\tau(M_{1}^k)$ is a displaced thermal state with variance $(2-\tau)/\tau$ and displaced by $\mu_k/\sqrt{\tau}$ from the origin in the phase space. Note that, unlike the original measurements, the noisy versions are not projective since the thermal states are not pure.

We use numerical methods to prove the incompatibility of these measurements after the loss channel. Consider a set of measurements that live in a $d^{\prime}$-dimensional Hilbert space---where $d^{\prime}$ can also be infinite. 
We then project these measurements into a $d$-dimensional subspace $d<d^{\prime}$. If the projected measurements are incompatible, then the original set should be incompatible too~\cite{Loulidi2021,PhysRevA.103.022203}. While this is not a necessary condition for proving incompatibility, it is sufficient and can be useful to deal with infinite-dimensional measurements. In fact, by truncating the Hilbert space into a finite-dimensional subspace, one can use the standard SDP formalism to test the incompatibility of measurements~\cite{Cavalcanti_2017,RevModPhys.95.011003}. We define the projector operators $\Pi_d = \sum_{k=0}^{d-1} \ket{k}\bra{k}$ in a subspace of the Hilbert space with maximum $d-1$ photons. Then, we investigate the incompatibility of the set $\left\{\{\Pi_d\,\mathcal{E}^*_\tau (M_{j}^k)\Pi_d\}_j\right\}_{k}$ over $d$-dimensional subspace using a SDP algorithm e.g., provided in Ref.~\cite{Cavalcanti_2017}. If we find that for some $d$ the set of truncated measurements is incompatible, then the original measurements are also incompatible. 

By using the $Q$ function~\eqref{eq:Q-function}, we can find the matrix elements of POVM elements $\mathcal{E}^*_\tau (M_{j}^k)$ in the Fock basis~\cite{Mandel_Wolf_1995} 
\begin{equation}\label{eq:Q-to-Fock}
    \bra{k}\mathcal{E}^*_\tau (M_{j}^k)\ket{j}=\frac{\pi}{\sqrt{k! j!}} \partial_{\alpha}^j
 \partial_{{\alpha}^*}^{k} \!\bigl[ e^{|\alpha|^2} \! Q\big(\alpha|\mathcal{E}^*_\tau(M_{1}^k)\bigr]\!\Big|_{\alpha=0}\!,
\end{equation}
where $\partial_{\alpha}^j\coloneqq\partial^j/\partial\alpha^j$ and $\alpha$ and
its complex conjugate $\alpha^*$ are treated as independent variables. This expression can be verified by expanding the coherent state $\ket{\alpha}$ in the $Q$ function in the Fock basis. Using Eq.~\eqnref{eq:Q-to-Fock}, we can find projected measurement operators onto the truncated Hilbert space for $d=3$, 
\begin{gather}
	\Pi_3\,\mathcal{E}_{\tau}^* (\ket{\mu_k}\bra{\mu_k})\Pi_3 
	= e^{\frac{-|\mu_k|^2 }{2}} 
	\begin{psmallmatrix}
		1 &    \frac{\sqrt{\tau}\mu_k^*}{\sqrt{2}} & \frac{\tau[\mu_k^*]^2}{\sqrt{2}}\\
		 \frac{\sqrt{\tau}\mu_k}{\sqrt{2}} & A_{11} & A_{21}^*\\
		\frac{\tau[\mu_k]^2}{\sqrt{2}} & A_{21} & A_{22} 
	\end{psmallmatrix}
	,\nonumber\\
     \Pi_3\, \mathcal{E}_{\tau}^* (I-\ket{\mu_k}\bra{\mu_k})\Pi_3 = I_3 - \Pi_3 \mathcal{E}_{\tau}^* (\ket{\mu_k}\bra{\mu_k})\Pi_3, \nonumber
     \label{eq:P_2_coherent}
\end{gather}
where $I_3$ is the identity matrix in dimension $d=3$, and
\begin{small}
\begin{align}
    A_{11} & \coloneqq \frac12\tau|\mu_k|^2 - \tau+1,\nonumber\\
    A_{21} & \coloneqq\frac12 \sqrt{\tau}\mu_k e^{-\frac{|\mu_k|^2}{2}} \big(\tau(|\mu_k|^2-4)+2\big),\nonumber\\
    A_{22} & \coloneqq\frac18 e^{-\frac{|\mu_k|^2}{2}} \big(\tau|\mu_k|^2[ \tau|\mu_k|^2 + 8(1-\tau)]+ 8(1-\tau)^2 \big) \nonumber
\end{align}
\end{small}

Having these at hand we can numerically confirm that our $n+1$ proposed measurement set given by Eq.~\eqref{eq:disp_photo_det} remain incompatible under $\tau=1/n$ even when truncated into $d=3$~~\footnote{Codes are publicly available on \href{https://github.com/Mehboudi/Incompatibility_under_pure_loss.git}{this GitHub repository}.}. Table~\ref{Table:r} summarises an example value of $r$ that guarantees the set to remain incompatible. 
Furthermore, we also numerically confirm that for $2\leq n \leq 13$, by setting $r=0.015$, our set of $n+1$ measurements remain incompatible for any $\tau \geq 1/n$ even when projected into $d=2$ (the qubit subspace). Note that any incompatible set of measurements can be used to steer a maximally entangled state~\cite{Quintino2014,Uola2014}. A maximally entangled state in the 2-dimensional subspace for each mode is $\ket{\psi_2}=(\ket{01}+\ket{10})/\sqrt{2}$, which can be prepared by injecting a single-photon to a 50:50 beam splitter. Therefore, our numerical results imply $\ket{\psi_2}$ remains steerable even when subjected to a loss with $\tau = 1/13$.

While the current challenge in considering higher losses is the computational power, as the problem size (i.e., the number of POVM elements of the parent measurement) increases exponentially with $n$, we conjecture that our proposed measurements remain incompatible for higher losses. This conjecture is further supported by the following result.

{\it Result 4.---} No loss channel breaks the incompatibility of all measurements. Consequently, loss channels cannot destroy the quantum steerability of all states.

{\it Proof of Result 4.---}Consider the unambiguous state discrimination (USD) task between $n$ symmeteric coherent states $\{\ket{r e^{i 2\pi t/n}} \}_{t=1}^{n}$ with $r\ge0$. Subjecting these states to a loss channel with transmissivity $\tau$ reduces their amplitudes by a factor of $\sqrt{\tau}$. After the loss channel, we perform the displaced on-off photodetection measurements~\eqnref{eq:disp_photo_det} with $\mu_k=r\sqrt{\tau} e^{i 2\pi k/n}$, and $k\in\{1,\dots, n\}$. Suppose that $\mathcal{E}_{\tau_b}$ is incompatibility breaking for all measurements with some strictly positive $\tau_b$. Hence, the displaced on-off photodetection measurements under loss with the POVM $\{\mathcal{E}^*_{\tau_b}(M_{1}^k), \mathcal{E}^*_{\tau_b}(M_{2}^k)\}_{k=1}^n$ would be jointly measurable. If the second outcome from the $k$th measurement with probability $\tr(\ket{r e^{i 2\pi t/n}}\bra{r e^{i 2\pi t/n}} \mathcal{E}^*_{\tau_b}(M_{2}^k)) $ is detected, then we can certainly say that $t\neq k$. Therefore, by performing all these measurements simultaneously (or in other words, by performing their parent measurement) and detecting $n-1$ of these outcomes, we can unambiguously identify the initial coherent state with probability $\prod_{k=1}^{n-1}(1-e^{-\tau_b r^2 |e^{i2\pi k/n}-1|^2})$. This probability for $r\ll1$ approximately becomes $n^2 r^{2(n-1)}\tau_b^{n-1}$, which is larger than the maximum achievable success probability~\cite{CHEFLES1998,vanEnk2002} if $n$ is chosen such that $n!>\tau_b^{1-n}$; see Appendix~\ref{app:USD}. 
This implies that measurements $\{\mathcal{E}^*_{\tau_b}(M_{1}^k), \mathcal{E}^*_{\tau_b}(M_{2}^k)\}_{k=1}^n$ cannot be jointly measurable. Therefore, there is no loss channel $\mathcal{E}_{\tau_b}$ that breaks the incompatibility of all measurements. Given the link between incompatibility and steering~\cite{PhysRevA.96.042331,Quintino2014,Uola2014}, the set $\{\mathcal{E}^*_{\tau}(M_{1}^k), \mathcal{E}^*_{\tau}(M_{2}^k)\}_{k=1}^n$ with $n! > \tau^{1-n}$ can be used for quantum steering. This means that there must be bipartite entangled states that remain steerable after the loss channel $\mathcal{E}_{\tau}$ applied to one subsystem.

{\it Conclusions.---}We 
have shown that under transmissivity $\tau \leq 1/n$ any set of $n$ measurements becomes compatible. Moreover, through constructive proof, we have also shown the existence of a set of $n+1$ measurements that remain incompatible under transmissivity $\tau \geq 1/n$. Our construction is based on measurements that are practically feasible within quantum optics platforms, as they require linear optics and on-off photodetection.  
We explicitly show this for $2\leq n \leq 13$.
However, we conjecture that the same is true for higher $n$. We back this conjecture by analytically showing that for any arbitrary low but strictly positive loss $\tau$ the set of $n$ displaced on-off photodetection measurements remains incompatible for some finite $n$ satisfying $n!>\tau^{1-n}$. 

Note that displaced on-off photodetection measurement can be viewed as a coarse-grained version of displaced number-resolving photodetection measurement, also known as unbalanced homodyne detection~\cite{Wallentowitz1996}. If a set of coarse-grained measurements are incompatible, the original set must also be incompatible. Therefore, our results also imply that unbalanced homodyne measurements remain incompatible under loss.

While continuous variable systems hold promise for many quantum information processing tasks, namely quantum communications, losses remain one of their main disadvantages and can limit the achievable distance between two ends of communications. 
Our results demonstrate that quantum steering remains achievable under pure loss, suggesting that one-sided device-independent protocols can be extended to longer distances with continuous-variable systems. An interesting direction for future research is to investigate whether our measurement set could be applied to studying Bell non-locality in continuous-variable systems under loss.

\begin{table}[t]
\vspace{.25cm}
\begin{minipage}{0.49\columnwidth}
\begin{tabular}{c|c|c}
	n & $r$ & $\tau_{\min}$ \\
	\hline
	2  & 0.005    & $1/n + 0.00005$ \\
	\hline
	3  & 0.010   & $1/n + 0.00018$\\
	\hline
	4  & 0.065    & $1/n + 0.00118$ \\
	\hline
	5  & 0.045    & $1/n + 0.00135$ \\
	\hline
 	6  & 0.035    & $1/n + 0.00100$  \\
	\hline
 \end{tabular}
 \end{minipage}
 \hspace{-.7cm}
 \begin{minipage}{0.49\columnwidth}
 \vspace{-.38cm}
 \begin{tabular}{c|c|c}
	n & $r$ & $\tau_{\min}$ \\
	\hline
	7  & 0.025    & $1/n + 0.00055$  \\
	\hline
	8  & 0.015    & $1/n + 0.00015$ \\
	\hline
	9  & 0.010    & $1/n + 0.00010$ \\
	\hline
	10  & 0.005    & $1/n + 0.00005$ \\
	\hline
\end{tabular}
\end{minipage}
 \caption{Feasible set of $n$ measurements that remain incompatible for transmissivity above $\tau > 1/n + \epsilon \eqqcolon \tau_{\min}$. These measurements are based on Eq.~\eqref{eq:disp_photo_det}, with $\mu_k = r\exp\left[i{\frac{2(k-1) \pi}{n+1}}\right]$ and with the $r$ parameter given by the corresponding column of the table.}
 \label{Table:r}
\end{table}

{\it Acknowledgements.---}We thank Roope Uola, Tobias Osborne, Florian Kanitschar, Salman Beigi,
and Ludovico Lami for sharing with us their insights on some mathematical and practical aspects of the problem.
The authors acknowledge
TU Wien Bibliothek for financial support through its Open Access Funding Programme. This research was funded in part by the Austrian Science Fund (FWF) [grant I 6047-N], and by the European Research Council (Consolidator grant ‘Cocoquest’ 101043705).
\bibliography{Refs}

\begin{thebibliography}{44}%
\makeatletter
\providecommand \@ifxundefined [1]{%
 \@ifx{#1\undefined}
}%
\providecommand \@ifnum [1]{%
 \ifnum #1\expandafter \@firstoftwo
 \else \expandafter \@secondoftwo
 \fi
}%
\providecommand \@ifx [1]{%
 \ifx #1\expandafter \@firstoftwo
 \else \expandafter \@secondoftwo
 \fi
}%
\providecommand \natexlab [1]{#1}%
\providecommand \enquote  [1]{``#1''}%
\providecommand \bibnamefont  [1]{#1}%
\providecommand \bibfnamefont [1]{#1}%
\providecommand \citenamefont [1]{#1}%
\providecommand \href@noop [0]{\@secondoftwo}%
\providecommand \href [0]{\begingroup \@sanitize@url \@href}%
\providecommand \@href[1]{\@@startlink{#1}\@@href}%
\providecommand \@@href[1]{\endgroup#1\@@endlink}%
\providecommand \@sanitize@url [0]{\catcode `\\12\catcode `\$12\catcode
  `\&12\catcode `\#12\catcode `\^12\catcode `\_12\catcode `\%12\relax}%
\providecommand \@@startlink[1]{}%
\providecommand \@@endlink[0]{}%
\providecommand \url  [0]{\begingroup\@sanitize@url \@url }%
\providecommand \@url [1]{\endgroup\@href {#1}{\urlprefix }}%
\providecommand \urlprefix  [0]{URL }%
\providecommand \Eprint [0]{\href }%
\providecommand \doibase [0]{http://dx.doi.org/}%
\providecommand \selectlanguage [0]{\@gobble}%
\providecommand \bibinfo  [0]{\@secondoftwo}%
\providecommand \bibfield  [0]{\@secondoftwo}%
\providecommand \translation [1]{[#1]}%
\providecommand \BibitemOpen [0]{}%
\providecommand \bibitemStop [0]{}%
\providecommand \bibitemNoStop [0]{.\EOS\space}%
\providecommand \EOS [0]{\spacefactor3000\relax}%
\providecommand \BibitemShut  [1]{\csname bibitem#1\endcsname}%
\let\auto@bib@innerbib\@empty
\bibitem [{\citenamefont {G\"uhne}\ \emph {et~al.}(2023)\citenamefont
  {G\"uhne}, \citenamefont {Haapasalo}, \citenamefont {Kraft}, \citenamefont
  {Pellonp\"a\"a},\ and\ \citenamefont {Uola}}]{RevModPhys.95.011003}%
  \BibitemOpen
  \bibfield  {author} {\bibinfo {author} {\bibfnamefont {O.}~\bibnamefont
  {G\"uhne}}, \bibinfo {author} {\bibfnamefont {E.}~\bibnamefont {Haapasalo}},
  \bibinfo {author} {\bibfnamefont {T.}~\bibnamefont {Kraft}}, \bibinfo
  {author} {\bibfnamefont {J.-P.}\ \bibnamefont {Pellonp\"a\"a}}, \ and\
  \bibinfo {author} {\bibfnamefont {R.}~\bibnamefont {Uola}},\ }\href {\doibase
  10.1103/RevModPhys.95.011003} {\enquote {\bibinfo {title} {Colloquium:
  Incompatible measurements in quantum information science},}\ } (\bibinfo
  {year} {2023})\BibitemShut {NoStop}%
\bibitem [{\citenamefont {Brunner}\ \emph {et~al.}(2014)\citenamefont
  {Brunner}, \citenamefont {Cavalcanti}, \citenamefont {Pironio}, \citenamefont
  {Scarani},\ and\ \citenamefont {Wehner}}]{RevModPhys.86.419}%
  \BibitemOpen
  \bibfield  {author} {\bibinfo {author} {\bibfnamefont {N.}~\bibnamefont
  {Brunner}}, \bibinfo {author} {\bibfnamefont {D.}~\bibnamefont {Cavalcanti}},
  \bibinfo {author} {\bibfnamefont {S.}~\bibnamefont {Pironio}}, \bibinfo
  {author} {\bibfnamefont {V.}~\bibnamefont {Scarani}}, \ and\ \bibinfo
  {author} {\bibfnamefont {S.}~\bibnamefont {Wehner}},\ }\href {\doibase
  10.1103/RevModPhys.86.419} {\bibfield  {journal} {\bibinfo  {journal} {Rev.
  Mod. Phys.}\ }\textbf {\bibinfo {volume} {86}},\ \bibinfo {pages} {419}
  (\bibinfo {year} {2014})}\BibitemShut {NoStop}%
\bibitem [{\citenamefont {Uola}\ \emph {et~al.}(2020)\citenamefont {Uola},
  \citenamefont {Costa}, \citenamefont {Nguyen},\ and\ \citenamefont
  {G\"uhne}}]{RevModPhys.92.015001}%
  \BibitemOpen
  \bibfield  {author} {\bibinfo {author} {\bibfnamefont {R.}~\bibnamefont
  {Uola}}, \bibinfo {author} {\bibfnamefont {A.~C.~S.}\ \bibnamefont {Costa}},
  \bibinfo {author} {\bibfnamefont {H.~C.}\ \bibnamefont {Nguyen}}, \ and\
  \bibinfo {author} {\bibfnamefont {O.}~\bibnamefont {G\"uhne}},\ }\href
  {\doibase 10.1103/RevModPhys.92.015001} {\bibfield  {journal} {\bibinfo
  {journal} {Rev. Mod. Phys.}\ }\textbf {\bibinfo {volume} {92}},\ \bibinfo
  {pages} {015001} (\bibinfo {year} {2020})}\BibitemShut {NoStop}%
\bibitem [{\citenamefont {Xu}\ \emph {et~al.}(2020)\citenamefont {Xu},
  \citenamefont {Ma}, \citenamefont {Zhang}, \citenamefont {Lo},\ and\
  \citenamefont {Pan}}]{RevModPhys.92.025002}%
  \BibitemOpen
  \bibfield  {author} {\bibinfo {author} {\bibfnamefont {F.}~\bibnamefont
  {Xu}}, \bibinfo {author} {\bibfnamefont {X.}~\bibnamefont {Ma}}, \bibinfo
  {author} {\bibfnamefont {Q.}~\bibnamefont {Zhang}}, \bibinfo {author}
  {\bibfnamefont {H.-K.}\ \bibnamefont {Lo}}, \ and\ \bibinfo {author}
  {\bibfnamefont {J.-W.}\ \bibnamefont {Pan}},\ }\href {\doibase
  10.1103/RevModPhys.92.025002} {\bibfield  {journal} {\bibinfo  {journal}
  {Rev. Mod. Phys.}\ }\textbf {\bibinfo {volume} {92}},\ \bibinfo {pages}
  {025002} (\bibinfo {year} {2020})}\BibitemShut {NoStop}%
\bibitem [{\citenamefont {Gisin}\ and\ \citenamefont {Thew}(2007)}]{Gisin2007}%
  \BibitemOpen
  \bibfield  {author} {\bibinfo {author} {\bibfnamefont {N.}~\bibnamefont
  {Gisin}}\ and\ \bibinfo {author} {\bibfnamefont {R.}~\bibnamefont {Thew}},\
  }\href {\doibase 10.1038/nphoton.2007.22} {\bibfield  {journal} {\bibinfo
  {journal} {Nature Photonics}\ }\textbf {\bibinfo {volume} {1}},\ \bibinfo
  {pages} {165–171} (\bibinfo {year} {2007})}\BibitemShut {NoStop}%
\bibitem [{\citenamefont {Heisenberg}(1925)}]{Heisenberg1925}%
  \BibitemOpen
  \bibfield  {author} {\bibinfo {author} {\bibfnamefont {W.}~\bibnamefont
  {Heisenberg}},\ }\href {\doibase 10.1007/BF01328377} {\bibfield  {journal}
  {\bibinfo  {journal} {Zeitschrift für Physik}\ }\textbf {\bibinfo {volume}
  {33}},\ \bibinfo {pages} {879} (\bibinfo {year} {1925})}\BibitemShut
  {NoStop}%
\bibitem [{\citenamefont {Robertson}(1929)}]{Robertson1929}%
  \BibitemOpen
  \bibfield  {author} {\bibinfo {author} {\bibfnamefont {H.~P.}\ \bibnamefont
  {Robertson}},\ }\href {\doibase 10.1103/PhysRev.34.163} {\bibfield  {journal}
  {\bibinfo  {journal} {Phys. Rev.}\ }\textbf {\bibinfo {volume} {34}},\
  \bibinfo {pages} {163} (\bibinfo {year} {1929})}\BibitemShut {NoStop}%
\bibitem [{\citenamefont {Horodecki}\ \emph {et~al.}(2009)\citenamefont
  {Horodecki}, \citenamefont {Horodecki}, \citenamefont {Horodecki},\ and\
  \citenamefont {Horodecki}}]{Horodecki2009}%
  \BibitemOpen
  \bibfield  {author} {\bibinfo {author} {\bibfnamefont {R.}~\bibnamefont
  {Horodecki}}, \bibinfo {author} {\bibfnamefont {P.}~\bibnamefont
  {Horodecki}}, \bibinfo {author} {\bibfnamefont {M.}~\bibnamefont
  {Horodecki}}, \ and\ \bibinfo {author} {\bibfnamefont {K.}~\bibnamefont
  {Horodecki}},\ }\href {\doibase 10.1103/RevModPhys.81.865} {\bibfield
  {journal} {\bibinfo  {journal} {Rev. Mod. Phys.}\ }\textbf {\bibinfo {volume}
  {81}},\ \bibinfo {pages} {865} (\bibinfo {year} {2009})}\BibitemShut
  {NoStop}%
\bibitem [{\citenamefont {Quintino}\ \emph {et~al.}(2014)\citenamefont
  {Quintino}, \citenamefont {V\'ertesi},\ and\ \citenamefont
  {Brunner}}]{Quintino2014}%
  \BibitemOpen
  \bibfield  {author} {\bibinfo {author} {\bibfnamefont {M.~T.}\ \bibnamefont
  {Quintino}}, \bibinfo {author} {\bibfnamefont {T.}~\bibnamefont {V\'ertesi}},
  \ and\ \bibinfo {author} {\bibfnamefont {N.}~\bibnamefont {Brunner}},\ }\href
  {\doibase 10.1103/PhysRevLett.113.160402} {\bibfield  {journal} {\bibinfo
  {journal} {Phys. Rev. Lett.}\ }\textbf {\bibinfo {volume} {113}},\ \bibinfo
  {pages} {160402} (\bibinfo {year} {2014})}\BibitemShut {NoStop}%
\bibitem [{\citenamefont {Uola}\ \emph {et~al.}(2014)\citenamefont {Uola},
  \citenamefont {Moroder},\ and\ \citenamefont {G\"uhne}}]{Uola2014}%
  \BibitemOpen
  \bibfield  {author} {\bibinfo {author} {\bibfnamefont {R.}~\bibnamefont
  {Uola}}, \bibinfo {author} {\bibfnamefont {T.}~\bibnamefont {Moroder}}, \
  and\ \bibinfo {author} {\bibfnamefont {O.}~\bibnamefont {G\"uhne}},\ }\href
  {\doibase 10.1103/PhysRevLett.113.160403} {\bibfield  {journal} {\bibinfo
  {journal} {Phys. Rev. Lett.}\ }\textbf {\bibinfo {volume} {113}},\ \bibinfo
  {pages} {160403} (\bibinfo {year} {2014})}\BibitemShut {NoStop}%
\bibitem [{\citenamefont {Skrzypczyk}\ \emph {et~al.}(2019)\citenamefont
  {Skrzypczyk}, \citenamefont {\ifmmode \check{S}\else
  \v{S}\fi{}upi\ifmmode~\acute{c}\else \'{c}\fi{}},\ and\ \citenamefont
  {Cavalcanti}}]{Skrzypczyk2019QSD}%
  \BibitemOpen
  \bibfield  {author} {\bibinfo {author} {\bibfnamefont {P.}~\bibnamefont
  {Skrzypczyk}}, \bibinfo {author} {\bibfnamefont {I.}~\bibnamefont {\ifmmode
  \check{S}\else \v{S}\fi{}upi\ifmmode~\acute{c}\else \'{c}\fi{}}}, \ and\
  \bibinfo {author} {\bibfnamefont {D.}~\bibnamefont {Cavalcanti}},\ }\href
  {\doibase 10.1103/PhysRevLett.122.130403} {\bibfield  {journal} {\bibinfo
  {journal} {Phys. Rev. Lett.}\ }\textbf {\bibinfo {volume} {122}},\ \bibinfo
  {pages} {130403} (\bibinfo {year} {2019})}\BibitemShut {NoStop}%
\bibitem [{\citenamefont {Carmeli}\ \emph {et~al.}(2019)\citenamefont
  {Carmeli}, \citenamefont {Heinosaari},\ and\ \citenamefont
  {Toigo}}]{Carmeli2019QSD}%
  \BibitemOpen
  \bibfield  {author} {\bibinfo {author} {\bibfnamefont {C.}~\bibnamefont
  {Carmeli}}, \bibinfo {author} {\bibfnamefont {T.}~\bibnamefont {Heinosaari}},
  \ and\ \bibinfo {author} {\bibfnamefont {A.}~\bibnamefont {Toigo}},\ }\href
  {\doibase 10.1103/PhysRevLett.122.130402} {\bibfield  {journal} {\bibinfo
  {journal} {Phys. Rev. Lett.}\ }\textbf {\bibinfo {volume} {122}},\ \bibinfo
  {pages} {130402} (\bibinfo {year} {2019})}\BibitemShut {NoStop}%
\bibitem [{\citenamefont {Sen}\ \emph {et~al.}(2024)\citenamefont {Sen},
  \citenamefont {Halder},\ and\ \citenamefont {Sen}}]{Sen2024QSD}%
  \BibitemOpen
  \bibfield  {author} {\bibinfo {author} {\bibfnamefont {K.}~\bibnamefont
  {Sen}}, \bibinfo {author} {\bibfnamefont {S.}~\bibnamefont {Halder}}, \ and\
  \bibinfo {author} {\bibfnamefont {U.}~\bibnamefont {Sen}},\ }\href {\doibase
  10.1103/PhysRevA.109.012415} {\bibfield  {journal} {\bibinfo  {journal}
  {Phys. Rev. A}\ }\textbf {\bibinfo {volume} {109}},\ \bibinfo {pages}
  {012415} (\bibinfo {year} {2024})}\BibitemShut {NoStop}%
\bibitem [{\citenamefont {Saha}\ \emph {et~al.}(2023)\citenamefont {Saha},
  \citenamefont {Das}, \citenamefont {Das}, \citenamefont {Bhattacharya},\ and\
  \citenamefont {Majumdar}}]{Saha2023communication}%
  \BibitemOpen
  \bibfield  {author} {\bibinfo {author} {\bibfnamefont {D.}~\bibnamefont
  {Saha}}, \bibinfo {author} {\bibfnamefont {D.}~\bibnamefont {Das}}, \bibinfo
  {author} {\bibfnamefont {A.~K.}\ \bibnamefont {Das}}, \bibinfo {author}
  {\bibfnamefont {B.}~\bibnamefont {Bhattacharya}}, \ and\ \bibinfo {author}
  {\bibfnamefont {A.~S.}\ \bibnamefont {Majumdar}},\ }\href {\doibase
  10.1103/PhysRevA.107.062210} {\bibfield  {journal} {\bibinfo  {journal}
  {Phys. Rev. A}\ }\textbf {\bibinfo {volume} {107}},\ \bibinfo {pages}
  {062210} (\bibinfo {year} {2023})}\BibitemShut {NoStop}%
\bibitem [{\citenamefont {Gisin}\ \emph {et~al.}(2002)\citenamefont {Gisin},
  \citenamefont {Ribordy}, \citenamefont {Tittel},\ and\ \citenamefont
  {Zbinden}}]{Gisin2002}%
  \BibitemOpen
  \bibfield  {author} {\bibinfo {author} {\bibfnamefont {N.}~\bibnamefont
  {Gisin}}, \bibinfo {author} {\bibfnamefont {G.}~\bibnamefont {Ribordy}},
  \bibinfo {author} {\bibfnamefont {W.}~\bibnamefont {Tittel}}, \ and\ \bibinfo
  {author} {\bibfnamefont {H.}~\bibnamefont {Zbinden}},\ }\href {\doibase
  10.1103/RevModPhys.74.145} {\bibfield  {journal} {\bibinfo  {journal} {Rev.
  Mod. Phys.}\ }\textbf {\bibinfo {volume} {74}},\ \bibinfo {pages} {145}
  (\bibinfo {year} {2002})}\BibitemShut {NoStop}%
\bibitem [{\citenamefont {Paris}\ and\ \citenamefont
  {Rehacek}(2004)}]{paris2004quantum}%
  \BibitemOpen
  \bibfield  {author} {\bibinfo {author} {\bibfnamefont {M.}~\bibnamefont
  {Paris}}\ and\ \bibinfo {author} {\bibfnamefont {J.}~\bibnamefont
  {Rehacek}},\ }\href@noop {} {\emph {\bibinfo {title} {Quantum state
  estimation}}},\ Vol.\ \bibinfo {volume} {649}\ (\bibinfo  {publisher}
  {Springer Science \& Business Media},\ \bibinfo {year} {2004})\BibitemShut
  {NoStop}%
\bibitem [{\citenamefont {Zhu}(2022)}]{PRXQuantum2022}%
  \BibitemOpen
  \bibfield  {author} {\bibinfo {author} {\bibfnamefont {H.}~\bibnamefont
  {Zhu}},\ }\href {\doibase 10.1103/PRXQuantum.3.030306} {\bibfield  {journal}
  {\bibinfo  {journal} {PRX Quantum}\ }\textbf {\bibinfo {volume} {3}},\
  \bibinfo {pages} {030306} (\bibinfo {year} {2022})}\BibitemShut {NoStop}%
\bibitem [{\citenamefont {Skrzypczyk}\ and\ \citenamefont
  {Cavalcanti}(2015)}]{Skrzypczyk2015oneNth}%
  \BibitemOpen
  \bibfield  {author} {\bibinfo {author} {\bibfnamefont {P.}~\bibnamefont
  {Skrzypczyk}}\ and\ \bibinfo {author} {\bibfnamefont {D.}~\bibnamefont
  {Cavalcanti}},\ }\href {\doibase 10.1103/PhysRevA.92.022354} {\bibfield
  {journal} {\bibinfo  {journal} {Phys. Rev. A}\ }\textbf {\bibinfo {volume}
  {92}},\ \bibinfo {pages} {022354} (\bibinfo {year} {2015})}\BibitemShut
  {NoStop}%
\bibitem [{\citenamefont {Heinosaari}\ \emph
  {et~al.}(2015{\natexlab{a}})\citenamefont {Heinosaari}, \citenamefont
  {Kiukas},\ and\ \citenamefont {Reitzner}}]{Heinosaari2015Noise}%
  \BibitemOpen
  \bibfield  {author} {\bibinfo {author} {\bibfnamefont {T.}~\bibnamefont
  {Heinosaari}}, \bibinfo {author} {\bibfnamefont {J.}~\bibnamefont {Kiukas}},
  \ and\ \bibinfo {author} {\bibfnamefont {D.}~\bibnamefont {Reitzner}},\
  }\href {\doibase 10.1103/PhysRevA.92.022115} {\bibfield  {journal} {\bibinfo
  {journal} {Phys. Rev. A}\ }\textbf {\bibinfo {volume} {92}},\ \bibinfo
  {pages} {022115} (\bibinfo {year} {2015}{\natexlab{a}})}\BibitemShut
  {NoStop}%
\bibitem [{\citenamefont {Uola}\ \emph {et~al.}(2016)\citenamefont {Uola},
  \citenamefont {Luoma}, \citenamefont {Moroder},\ and\ \citenamefont
  {Heinosaari}}]{Uola2016JM}%
  \BibitemOpen
  \bibfield  {author} {\bibinfo {author} {\bibfnamefont {R.}~\bibnamefont
  {Uola}}, \bibinfo {author} {\bibfnamefont {K.}~\bibnamefont {Luoma}},
  \bibinfo {author} {\bibfnamefont {T.}~\bibnamefont {Moroder}}, \ and\
  \bibinfo {author} {\bibfnamefont {T.}~\bibnamefont {Heinosaari}},\ }\href
  {\doibase 10.1103/PhysRevA.94.022109} {\bibfield  {journal} {\bibinfo
  {journal} {Phys. Rev. A}\ }\textbf {\bibinfo {volume} {94}},\ \bibinfo
  {pages} {022109} (\bibinfo {year} {2016})}\BibitemShut {NoStop}%
\bibitem [{\citenamefont {Bavaresco}\ \emph {et~al.}(2017)\citenamefont
  {Bavaresco}, \citenamefont {Quintino}, \citenamefont {Guerini}, \citenamefont
  {Maciel}, \citenamefont {Cavalcanti},\ and\ \citenamefont
  {Cunha}}]{Bavaresco2017}%
  \BibitemOpen
  \bibfield  {author} {\bibinfo {author} {\bibfnamefont {J.}~\bibnamefont
  {Bavaresco}}, \bibinfo {author} {\bibfnamefont {M.~T.}\ \bibnamefont
  {Quintino}}, \bibinfo {author} {\bibfnamefont {L.}~\bibnamefont {Guerini}},
  \bibinfo {author} {\bibfnamefont {T.~O.}\ \bibnamefont {Maciel}}, \bibinfo
  {author} {\bibfnamefont {D.}~\bibnamefont {Cavalcanti}}, \ and\ \bibinfo
  {author} {\bibfnamefont {M.~T.}\ \bibnamefont {Cunha}},\ }\href {\doibase
  10.1103/PhysRevA.96.022110} {\bibfield  {journal} {\bibinfo  {journal} {Phys.
  Rev. A}\ }\textbf {\bibinfo {volume} {96}},\ \bibinfo {pages} {022110}
  (\bibinfo {year} {2017})}\BibitemShut {NoStop}%
\bibitem [{\citenamefont {Designolle}\ \emph {et~al.}(2019)\citenamefont
  {Designolle}, \citenamefont {Skrzypczyk}, \citenamefont {Fr\"owis},\ and\
  \citenamefont {Brunner}}]{Designolle2019Quantifying}%
  \BibitemOpen
  \bibfield  {author} {\bibinfo {author} {\bibfnamefont {S.}~\bibnamefont
  {Designolle}}, \bibinfo {author} {\bibfnamefont {P.}~\bibnamefont
  {Skrzypczyk}}, \bibinfo {author} {\bibfnamefont {F.}~\bibnamefont
  {Fr\"owis}}, \ and\ \bibinfo {author} {\bibfnamefont {N.}~\bibnamefont
  {Brunner}},\ }\href {\doibase 10.1103/PhysRevLett.122.050402} {\bibfield
  {journal} {\bibinfo  {journal} {Phys. Rev. Lett.}\ }\textbf {\bibinfo
  {volume} {122}},\ \bibinfo {pages} {050402} (\bibinfo {year}
  {2019})}\BibitemShut {NoStop}%
\bibitem [{\citenamefont {Uola}\ \emph
  {et~al.}(2021{\natexlab{a}})\citenamefont {Uola}, \citenamefont {Kraft},
  \citenamefont {Designolle}, \citenamefont {Miklin}, \citenamefont {Tavakoli},
  \citenamefont {Pellonp\"a\"a}, \citenamefont {G\"uhne},\ and\ \citenamefont
  {Brunner}}]{Uola2021subspace}%
  \BibitemOpen
  \bibfield  {author} {\bibinfo {author} {\bibfnamefont {R.}~\bibnamefont
  {Uola}}, \bibinfo {author} {\bibfnamefont {T.}~\bibnamefont {Kraft}},
  \bibinfo {author} {\bibfnamefont {S.}~\bibnamefont {Designolle}}, \bibinfo
  {author} {\bibfnamefont {N.}~\bibnamefont {Miklin}}, \bibinfo {author}
  {\bibfnamefont {A.}~\bibnamefont {Tavakoli}}, \bibinfo {author}
  {\bibfnamefont {J.-P.}\ \bibnamefont {Pellonp\"a\"a}}, \bibinfo {author}
  {\bibfnamefont {O.}~\bibnamefont {G\"uhne}}, \ and\ \bibinfo {author}
  {\bibfnamefont {N.}~\bibnamefont {Brunner}},\ }\href {\doibase
  10.1103/PhysRevA.103.022203} {\bibfield  {journal} {\bibinfo  {journal}
  {Phys. Rev. A}\ }\textbf {\bibinfo {volume} {103}},\ \bibinfo {pages}
  {022203} (\bibinfo {year} {2021}{\natexlab{a}})}\BibitemShut {NoStop}%
\bibitem [{\citenamefont {Jones}\ \emph {et~al.}(2023)\citenamefont {Jones},
  \citenamefont {Uola}, \citenamefont {Cope}, \citenamefont {Ioannou},
  \citenamefont {Designolle}, \citenamefont {Sekatski},\ and\ \citenamefont
  {Brunner}}]{Jones2023HighD}%
  \BibitemOpen
  \bibfield  {author} {\bibinfo {author} {\bibfnamefont {B.~D.~M.}\
  \bibnamefont {Jones}}, \bibinfo {author} {\bibfnamefont {R.}~\bibnamefont
  {Uola}}, \bibinfo {author} {\bibfnamefont {T.}~\bibnamefont {Cope}}, \bibinfo
  {author} {\bibfnamefont {M.}~\bibnamefont {Ioannou}}, \bibinfo {author}
  {\bibfnamefont {S.}~\bibnamefont {Designolle}}, \bibinfo {author}
  {\bibfnamefont {P.}~\bibnamefont {Sekatski}}, \ and\ \bibinfo {author}
  {\bibfnamefont {N.}~\bibnamefont {Brunner}},\ }\href {\doibase
  10.1103/PhysRevA.107.052425} {\bibfield  {journal} {\bibinfo  {journal}
  {Phys. Rev. A}\ }\textbf {\bibinfo {volume} {107}},\ \bibinfo {pages}
  {052425} (\bibinfo {year} {2023})}\BibitemShut {NoStop}%
\bibitem [{\citenamefont {Skrzypczyk}\ and\ \citenamefont
  {Cavalcanti}(2023)}]{skrzypczyk2023SDP}%
  \BibitemOpen
  \bibfield  {author} {\bibinfo {author} {\bibfnamefont {P.}~\bibnamefont
  {Skrzypczyk}}\ and\ \bibinfo {author} {\bibfnamefont {D.}~\bibnamefont
  {Cavalcanti}},\ }\href@noop {} {\bibfield  {journal} {\bibinfo  {journal}
  {arXiv preprint arXiv:2306.11637}\ } (\bibinfo {year} {2023})}\BibitemShut
  {NoStop}%
\bibitem [{\citenamefont {Heinosaari}\ \emph
  {et~al.}(2015{\natexlab{b}})\citenamefont {Heinosaari}, \citenamefont
  {Kiukas},\ and\ \citenamefont {Schultz}}]{Heinosaari2015Gaussian}%
  \BibitemOpen
  \bibfield  {author} {\bibinfo {author} {\bibfnamefont {T.}~\bibnamefont
  {Heinosaari}}, \bibinfo {author} {\bibfnamefont {J.}~\bibnamefont {Kiukas}},
  \ and\ \bibinfo {author} {\bibfnamefont {J.}~\bibnamefont {Schultz}},\ }\href
  {\doibase 10.1063/1.4928044} {\bibfield  {journal} {\bibinfo  {journal}
  {Journal of Mathematical Physics}\ }\textbf {\bibinfo {volume} {56}},\
  \bibinfo {pages} {082202} (\bibinfo {year} {2015}{\natexlab{b}})},\ \Eprint
  {http://arxiv.org/abs/https://pubs.aip.org/aip/jmp/article-pdf/doi/10.1063/1.4928044/15805120/082202\_1\_online.pdf}
  {https://pubs.aip.org/aip/jmp/article-pdf/doi/10.1063/1.4928044/15805120/082202\_1\_online.pdf}
  \BibitemShut {NoStop}%
\bibitem [{\citenamefont {Rahimi-Keshari}\ \emph {et~al.}(2021)\citenamefont
  {Rahimi-Keshari}, \citenamefont {Mehboudi}, \citenamefont {De~Santis},
  \citenamefont {Cavalcanti},\ and\ \citenamefont
  {Ac\'{\i}n}}]{PhysRevA.104.042212}%
  \BibitemOpen
  \bibfield  {author} {\bibinfo {author} {\bibfnamefont {S.}~\bibnamefont
  {Rahimi-Keshari}}, \bibinfo {author} {\bibfnamefont {M.}~\bibnamefont
  {Mehboudi}}, \bibinfo {author} {\bibfnamefont {D.}~\bibnamefont {De~Santis}},
  \bibinfo {author} {\bibfnamefont {D.}~\bibnamefont {Cavalcanti}}, \ and\
  \bibinfo {author} {\bibfnamefont {A.}~\bibnamefont {Ac\'{\i}n}},\ }\href
  {\doibase 10.1103/PhysRevA.104.042212} {\bibfield  {journal} {\bibinfo
  {journal} {Phys. Rev. A}\ }\textbf {\bibinfo {volume} {104}},\ \bibinfo
  {pages} {042212} (\bibinfo {year} {2021})}\BibitemShut {NoStop}%
\bibitem [{\citenamefont {Ji}\ \emph {et~al.}(2016)\citenamefont {Ji},
  \citenamefont {Lee}, \citenamefont {Park},\ and\ \citenamefont
  {Nha}}]{Ji2016steering}%
  \BibitemOpen
  \bibfield  {author} {\bibinfo {author} {\bibfnamefont {S.-W.}\ \bibnamefont
  {Ji}}, \bibinfo {author} {\bibfnamefont {J.}~\bibnamefont {Lee}}, \bibinfo
  {author} {\bibfnamefont {J.}~\bibnamefont {Park}}, \ and\ \bibinfo {author}
  {\bibfnamefont {H.}~\bibnamefont {Nha}},\ }\href {\doibase 10.1038/srep29729}
  {\bibfield  {journal} {\bibinfo  {journal} {Scientific Reports}\ }\textbf
  {\bibinfo {volume} {6}},\ \bibinfo {pages} {29729} (\bibinfo {year}
  {2016})}\BibitemShut {NoStop}%
\bibitem [{\citenamefont {Quintino}\ \emph {et~al.}(2016)\citenamefont
  {Quintino}, \citenamefont {Bowles}, \citenamefont {Hirsch},\ and\
  \citenamefont {Brunner}}]{PhysRevA.93.052115}%
  \BibitemOpen
  \bibfield  {author} {\bibinfo {author} {\bibfnamefont {M.~T.}\ \bibnamefont
  {Quintino}}, \bibinfo {author} {\bibfnamefont {J.}~\bibnamefont {Bowles}},
  \bibinfo {author} {\bibfnamefont {F.}~\bibnamefont {Hirsch}}, \ and\ \bibinfo
  {author} {\bibfnamefont {N.}~\bibnamefont {Brunner}},\ }\href {\doibase
  10.1103/PhysRevA.93.052115} {\bibfield  {journal} {\bibinfo  {journal} {Phys.
  Rev. A}\ }\textbf {\bibinfo {volume} {93}},\ \bibinfo {pages} {052115}
  (\bibinfo {year} {2016})}\BibitemShut {NoStop}%
\bibitem [{\citenamefont {Cavalcanti}\ and\ \citenamefont
  {Skrzypczyk}(2016{\natexlab{a}})}]{PhysRevA.93.052112}%
  \BibitemOpen
  \bibfield  {author} {\bibinfo {author} {\bibfnamefont {D.}~\bibnamefont
  {Cavalcanti}}\ and\ \bibinfo {author} {\bibfnamefont {P.}~\bibnamefont
  {Skrzypczyk}},\ }\href {\doibase 10.1103/PhysRevA.93.052112} {\bibfield
  {journal} {\bibinfo  {journal} {Phys. Rev. A}\ }\textbf {\bibinfo {volume}
  {93}},\ \bibinfo {pages} {052112} (\bibinfo {year}
  {2016}{\natexlab{a}})}\BibitemShut {NoStop}%
\bibitem [{\citenamefont {K\"uhn}\ and\ \citenamefont
  {Vogel}(2018)}]{Kuhn2018}%
  \BibitemOpen
  \bibfield  {author} {\bibinfo {author} {\bibfnamefont {B.}~\bibnamefont
  {K\"uhn}}\ and\ \bibinfo {author} {\bibfnamefont {W.}~\bibnamefont {Vogel}},\
  }\href {\doibase 10.1103/PhysRevA.98.053807} {\bibfield  {journal} {\bibinfo
  {journal} {Phys. Rev. A}\ }\textbf {\bibinfo {volume} {98}},\ \bibinfo
  {pages} {053807} (\bibinfo {year} {2018})}\BibitemShut {NoStop}%
\bibitem [{\citenamefont {Husimi}(1940)}]{Husimi}%
  \BibitemOpen
  \bibfield  {author} {\bibinfo {author} {\bibfnamefont {K.}~\bibnamefont
  {Husimi}},\ }\href {\doibase 10.11429/ppmsj1919.22.4_264} {\bibfield
  {journal} {\bibinfo  {journal} {Proceedings of the Physico-Mathematical
  Society of Japan. 3rd Series}\ }\textbf {\bibinfo {volume} {22}},\ \bibinfo
  {pages} {264} (\bibinfo {year} {1940})}\BibitemShut {NoStop}%
\bibitem [{\citenamefont {Cahill}\ and\ \citenamefont
  {Glauber}(1969)}]{Cahill-Glauber1969}%
  \BibitemOpen
  \bibfield  {author} {\bibinfo {author} {\bibfnamefont {K.~E.}\ \bibnamefont
  {Cahill}}\ and\ \bibinfo {author} {\bibfnamefont {R.~J.}\ \bibnamefont
  {Glauber}},\ }\href {\doibase 10.1103/PhysRev.177.1882} {\bibfield  {journal}
  {\bibinfo  {journal} {Phys. Rev.}\ }\textbf {\bibinfo {volume} {177}},\
  \bibinfo {pages} {1882} (\bibinfo {year} {1969})}\BibitemShut {NoStop}%
\bibitem [{\citenamefont {Loulidi}\ and\ \citenamefont
  {Nechita}(2021)}]{Loulidi2021}%
  \BibitemOpen
  \bibfield  {author} {\bibinfo {author} {\bibfnamefont {F.}~\bibnamefont
  {Loulidi}}\ and\ \bibinfo {author} {\bibfnamefont {I.}~\bibnamefont
  {Nechita}},\ }\href {\doibase 10.1063/5.0028658} {\bibfield  {journal}
  {\bibinfo  {journal} {Journal of Mathematical Physics}\ }\textbf {\bibinfo
  {volume} {62}} (\bibinfo {year} {2021}),\ 10.1063/5.0028658}\BibitemShut
  {NoStop}%
\bibitem [{\citenamefont {Uola}\ \emph
  {et~al.}(2021{\natexlab{b}})\citenamefont {Uola}, \citenamefont {Kraft},
  \citenamefont {Designolle}, \citenamefont {Miklin}, \citenamefont {Tavakoli},
  \citenamefont {Pellonp\"a\"a}, \citenamefont {G\"uhne},\ and\ \citenamefont
  {Brunner}}]{PhysRevA.103.022203}%
  \BibitemOpen
  \bibfield  {author} {\bibinfo {author} {\bibfnamefont {R.}~\bibnamefont
  {Uola}}, \bibinfo {author} {\bibfnamefont {T.}~\bibnamefont {Kraft}},
  \bibinfo {author} {\bibfnamefont {S.}~\bibnamefont {Designolle}}, \bibinfo
  {author} {\bibfnamefont {N.}~\bibnamefont {Miklin}}, \bibinfo {author}
  {\bibfnamefont {A.}~\bibnamefont {Tavakoli}}, \bibinfo {author}
  {\bibfnamefont {J.-P.}\ \bibnamefont {Pellonp\"a\"a}}, \bibinfo {author}
  {\bibfnamefont {O.}~\bibnamefont {G\"uhne}}, \ and\ \bibinfo {author}
  {\bibfnamefont {N.}~\bibnamefont {Brunner}},\ }\href {\doibase
  10.1103/PhysRevA.103.022203} {\bibfield  {journal} {\bibinfo  {journal}
  {Phys. Rev. A}\ }\textbf {\bibinfo {volume} {103}},\ \bibinfo {pages}
  {022203} (\bibinfo {year} {2021}{\natexlab{b}})}\BibitemShut {NoStop}%
\bibitem [{\citenamefont {Cavalcanti}\ and\ \citenamefont
  {Skrzypczyk}(2016{\natexlab{b}})}]{Cavalcanti_2017}%
  \BibitemOpen
  \bibfield  {author} {\bibinfo {author} {\bibfnamefont {D.}~\bibnamefont
  {Cavalcanti}}\ and\ \bibinfo {author} {\bibfnamefont {P.}~\bibnamefont
  {Skrzypczyk}},\ }\href {\doibase 10.1088/1361-6633/80/2/024001} {\bibfield
  {journal} {\bibinfo  {journal} {Reports on Progress in Physics}\ }\textbf
  {\bibinfo {volume} {80}},\ \bibinfo {pages} {024001} (\bibinfo {year}
  {2016}{\natexlab{b}})}\BibitemShut {NoStop}%
\bibitem [{\citenamefont {Mandel}\ and\ \citenamefont
  {Wolf}(1995)}]{Mandel_Wolf_1995}%
  \BibitemOpen
  \bibfield  {author} {\bibinfo {author} {\bibfnamefont {L.}~\bibnamefont
  {Mandel}}\ and\ \bibinfo {author} {\bibfnamefont {E.}~\bibnamefont {Wolf}},\
  }\href@noop {} {\emph {\bibinfo {title} {Optical Coherence and Quantum
  Optics}}}\ (\bibinfo  {publisher} {Cambridge University Press},\ \bibinfo
  {year} {1995})\BibitemShut {NoStop}%
\bibitem [{Note1()}]{Note1}%
  \BibitemOpen
  \bibinfo {note} {Codes are publicly available on \protect \href
  {https://github.com/Mehboudi/Incompatibility_under_pure_loss.git}{this GitHub
  repository}.}\BibitemShut {Stop}%
\bibitem [{\citenamefont {Chefles}\ and\ \citenamefont
  {Barnett}(1998)}]{CHEFLES1998}%
  \BibitemOpen
  \bibfield  {author} {\bibinfo {author} {\bibfnamefont {A.}~\bibnamefont
  {Chefles}}\ and\ \bibinfo {author} {\bibfnamefont {S.~M.}\ \bibnamefont
  {Barnett}},\ }\href {\doibase https://doi.org/10.1016/S0375-9601(98)00827-5}
  {\bibfield  {journal} {\bibinfo  {journal} {Physics Letters A}\ }\textbf
  {\bibinfo {volume} {250}},\ \bibinfo {pages} {223} (\bibinfo {year}
  {1998})}\BibitemShut {NoStop}%
\bibitem [{\citenamefont {van Enk}(2002)}]{vanEnk2002}%
  \BibitemOpen
  \bibfield  {author} {\bibinfo {author} {\bibfnamefont {S.~J.}\ \bibnamefont
  {van Enk}},\ }\href {\doibase 10.1103/PhysRevA.66.042313} {\bibfield
  {journal} {\bibinfo  {journal} {Phys. Rev. A}\ }\textbf {\bibinfo {volume}
  {66}},\ \bibinfo {pages} {042313} (\bibinfo {year} {2002})}\BibitemShut
  {NoStop}%
\bibitem [{\citenamefont {Kiukas}\ \emph {et~al.}(2017)\citenamefont {Kiukas},
  \citenamefont {Budroni}, \citenamefont {Uola},\ and\ \citenamefont
  {Pellonp\"a\"a}}]{PhysRevA.96.042331}%
  \BibitemOpen
  \bibfield  {author} {\bibinfo {author} {\bibfnamefont {J.}~\bibnamefont
  {Kiukas}}, \bibinfo {author} {\bibfnamefont {C.}~\bibnamefont {Budroni}},
  \bibinfo {author} {\bibfnamefont {R.}~\bibnamefont {Uola}}, \ and\ \bibinfo
  {author} {\bibfnamefont {J.-P.}\ \bibnamefont {Pellonp\"a\"a}},\ }\href
  {\doibase 10.1103/PhysRevA.96.042331} {\bibfield  {journal} {\bibinfo
  {journal} {Phys. Rev. A}\ }\textbf {\bibinfo {volume} {96}},\ \bibinfo
  {pages} {042331} (\bibinfo {year} {2017})}\BibitemShut {NoStop}%
\bibitem [{\citenamefont {Wallentowitz}\ and\ \citenamefont
  {Vogel}(1996)}]{Wallentowitz1996}%
  \BibitemOpen
  \bibfield  {author} {\bibinfo {author} {\bibfnamefont {S.}~\bibnamefont
  {Wallentowitz}}\ and\ \bibinfo {author} {\bibfnamefont {W.}~\bibnamefont
  {Vogel}},\ }\href {\doibase 10.1103/PhysRevA.53.4528} {\bibfield  {journal}
  {\bibinfo  {journal} {Phys. Rev. A}\ }\textbf {\bibinfo {volume} {53}},\
  \bibinfo {pages} {4528} (\bibinfo {year} {1996})}\BibitemShut {NoStop}%
\bibitem [{\citenamefont {Yu}\ and\ \citenamefont {Oh}(2013)}]{yu2013quantum}%
  \BibitemOpen
  \bibfield  {author} {\bibinfo {author} {\bibfnamefont {S.}~\bibnamefont
  {Yu}}\ and\ \bibinfo {author} {\bibfnamefont {C.}~\bibnamefont {Oh}},\ }\href
  {\doibase https://doi.org/10.48550/arXiv.1312.6470} {\bibfield  {journal}
  {\bibinfo  {journal} {arXiv preprint arXiv:1312.6470}\ } (\bibinfo {year}
  {2013}),\ https://doi.org/10.48550/arXiv.1312.6470}\BibitemShut {NoStop}%
\bibitem [{\citenamefont {Yu}\ \emph {et~al.}(2010)\citenamefont {Yu},
  \citenamefont {Liu}, \citenamefont {Li},\ and\ \citenamefont
  {Oh}}]{PhysRevA.81.062116}%
  \BibitemOpen
  \bibfield  {author} {\bibinfo {author} {\bibfnamefont {S.}~\bibnamefont
  {Yu}}, \bibinfo {author} {\bibfnamefont {N.-l.}\ \bibnamefont {Liu}},
  \bibinfo {author} {\bibfnamefont {L.}~\bibnamefont {Li}}, \ and\ \bibinfo
  {author} {\bibfnamefont {C.~H.}\ \bibnamefont {Oh}},\ }\href {\doibase
  10.1103/PhysRevA.81.062116} {\bibfield  {journal} {\bibinfo  {journal} {Phys.
  Rev. A}\ }\textbf {\bibinfo {volume} {81}},\ \bibinfo {pages} {062116}
  (\bibinfo {year} {2010})}\BibitemShut {NoStop}%
\end{thebibliography}%
\appendix
\onecolumngrid

\section{Marginal of the parent measurement}\label{app:marginals}
In the main text, we claim that Eq.~\eqref{eq:parent_LON} gives the parent measurement ${\tilde M}_{\vec {\bm a}}$ for a set of $n$ measurements $\{\{{\mathcal E}_{1/n}^*(M_{\bm a_j}^j)\}_{\bm a_j}\}_{j=1}^{n}$. One could easily see that this is indeed the case, as $\sum_{\bm a_j\neq k} {\tilde M}_{\vec{\bm a}} = M_{{\bm a}_k}^k$. 
To confirm this, take an arbitrary state $\rho$, for which we have
\begin{align}
  {\rm tr}[\rho\sum_{\bm a_{j\neq k}} {\tilde M}_{\vec {\bm a}}] 
  &= {\rm tr}\Bigg[\rho\bra{0^{n-1}}\, \mathcal{U}_{\rm LON}^\dagger \Bigg(\! \sum_{\bm a_{j\neq k}} M_{{\bm a}_1}^1\otimes \cdots \otimes M_{{\bm a}_n}^n\!\Bigg)  \mathcal{U}_{\rm LON}\ket{0^{n-1}}\Bigg]\nonumber\\
  & = {\rm tr}\Big[\big(\rho\otimes\ket{0^{n-1}}\bra{0^{n-1}}\big)\, \mathcal{U}_{\rm LON}^{\dagger} \big(I^{\otimes k-1} \otimes M_{{\bm a}_k}^k \otimes I^{\otimes n-k}\big)\, \mathcal{U}_{\rm LON}\Big]\nonumber\\
  & = {\rm tr}\Big[\mathcal{U}_{\rm LON}\big(\rho\otimes\ket{0^{n-1}}\bra{0^{n-1}}\big)\, \mathcal{U}_{\rm LON}^{\dagger} \big(I^{\otimes k-1} \otimes M_{{\bm a}_k}^k \otimes I^{\otimes n-k}\big) \Big]\nonumber\\
  & = {\rm tr}_k\Big[{\rm tr}_{k^c}\!\Big[{\cal E}^{(n)}_{\rm LON}(\rho) \Big] M_{{\bm a}_k}^k \Big]\nonumber\\
  & = {\rm tr}_k\left[{\cal E}_{\frac{1}{n}}(\rho) M_{{\bm a}_k}^k\right] =  {\rm tr}_k\left[\rho\, {\cal E}^*_{\frac{1}{n}}(M_{\bm a_k}^k)\right],
\end{align}
where in the fourth line we used the fact that a loss channel can be viewed as a single-mode marginal channel of an LON. Thus, we have $\sum_{\bm a_{j\neq k}} {\tilde M}_{\vec {\bm a}} = {\cal E}^*_{\frac{1}{n}}(M_{\bm a_k}^k)$.

\section{Any two displaced on-off photodetection measurements are incompatible}\label{app:2_disp}
Consider two measurements with POVM elements
\begin{align}
    M_{1}^j = \ket{\mu_j}\bra{\mu_j},\hspace{1cm} M_{2}^j = I - M_{1}^j.
\end{align}
These two measurements are incompatible unless $\mu_1 = \mu_2$. The proof is by contradiction. Suppose the two are compatible, therefore there exists a parent measurement with four POVM elements $\{M_{jk}\}_{\{j,k\}\in\{1,2\}^2}$. Take $\ket{\mu_1^{\perp}}$ and $\ket{\mu_2^{\perp}}$---which are not coherent states---to be any states that are orthogonal to the coherent states $\ket{\mu_1}$ and $\ket{\mu_2}$, respectively. We should have
\begin{align}
    M_{11}+M_{12} & = \ket{\mu_1}\bra{\mu_1}\Rightarrow \bra{\mu_1^{\perp}}M_{11} + M_{12}\ket{\mu_1^{\perp}} = 0,\nonumber\\
    M_{11}+M_{21} & = \ket{\mu_2}\bra{\mu_2}\Rightarrow \bra{\mu_2^{\perp}}M_{11} + M_{21}\ket{\mu_2^{\perp}} = 0.
\end{align}
The above equations hold for \textit{any} $\ket{\mu_1^{\perp}}$ and $\ket{\mu_2^{\perp}}$. Thus, we should have
\begin{gather}
 M_{11} = a_{11}\ket{\mu_1}\bra{\mu_1},\hspace{.5 cm}M_{12} = a_{12}\ket{\mu_1}\bra{\mu_1}\nonumber\\
 M_{11} = b_{11}\ket{\mu_2}\bra{\mu_2},\hspace{.5 cm}M_{21} = b_{21}\ket{\mu_2}\bra{\mu_2}.
\end{gather}
This is only possible if $a_{11}=b_{11}=0$, and $a_{12}=b_{21}=1$.
On the other hand, we should have $\sum_{jk}M_{jk} = I$. Thus,
\begin{align}
    M_{22} = I - M_{12} - M_{21} - M_{11} = I - \ket{\mu_1}\bra{\mu_1} - \ket{\mu_2}\bra{\mu_2},
\end{align}
which is not a positive semi-definite operator (to see this note that e.g., $\bra{\mu_1} M_{22} \ket{\mu_1} = -|\braket{\mu_1}{\mu_2}|^2 < 0$). This is a contradiction, thus the two measurements are incompatible.
\section{Incompatibility of a pair of displaced on-off photodetection measurements under loss}\label{app:2_disp_tau}
Generally, we have no simple analytical solution for finding out whether a set of measurements in finite $d$ are incompatible. However, for sets of $2$ and $3$ qubit measurements, there are some techniques~\cite{RevModPhys.95.011003}. In particular, any two-outcome measurement for a qubit can be written as
\begin{align}
	A_{\pm}^i = \frac{1}{2}[(1\pm\gamma_i) \pm \vec m_i.\vec \sigma].
\end{align}
The parameter $\gamma_i$ represents the bias of the measurement. For the specific measurements that we consider one can simply check that this parameter is non-zero, i.e., the measurements are biased. Unfortunately, only for unbiased measurements, there exists a criterion for incompatibility of a set of any three measurements~\cite{yu2013quantum}. However, even for biased estimators, there exists a condition for incompatibility of any two measurements~\cite{PhysRevA.81.062116,RevModPhys.95.011003}, according to which, any two measurements are incompatible if and only if the following inequality is violated
\begin{align}
	{\rm Test}\coloneqq (1-F_1^2-F_2^2)\left(1-(\gamma_1/F_1)^2-(\gamma_2/F_2)^2\right)-(\vec m_1 . \vec m_2 - \gamma_1\gamma_2)^2 \leqslant 0,
\end{align}
with $F_i\coloneqq 1/2 [\sqrt{(1+\gamma_i)^2-\vec m_i.\vec m_i} + \sqrt{(1-\gamma_i)^2-\vec m_i.\vec m_i} ]$. For our specific choices of measurements, this reads
\begin{align}
	{\rm Test} = 16\tau (2\tau - 1)r^2 + {\cal O}(r^4),
\end{align}
which means by taking a small enough---but finite---$r$ the two measurements remain incompatible for any loss $\tau>1/2$. In other words, the two measurements are among the most robust against loss.

\section{Unambiguous state discrimination between symmetric coherent states}\label{app:USD}

Given $n$ symmetric coherent states $\{\ket{r e^{i 2\pi t/n}} \}_{t=1}^{n}$, where $r>0$, the bound on the maximum success probability to achieve unambiguous state discrimination (USD) is given by~\cite{CHEFLES1998}
\begin{align}
    P_{\rm D}^{(n)}= \min_{t=1,\cdots,n} \sum_{j=1}^{n} e^{i 2\pi j t/n} \exp\!\Big(r^2 \big(e^{i 2\pi j /n}-1\big) \Big).
\end{align}
For small values of $r$, this can be approximated by
\begin{equation}\label{eq:AppOptProb-USD}
    P_{\rm D}^{(n)}\approx \frac{n^2 r^{2(n-1)}}{n!}. 
\end{equation}

A method for USD~\cite{vanEnk2002} is to split input coherent states $\ket{r e^{i 2\pi t/n}}$ into $n$ coherent states with reduced amplitudes $\ket{ r e^{i 2\pi t/n}/\sqrt{n}}$, using a linear-optical network (LON) with a transfer matrix whose first row is $U_{1j}=1/\sqrt{n}$ for $j=1,\dots,n$ (see Fig.~\ref{fig:LOSS_BS}). Then the displaced on-off photodetection measurement with $\{M_{1}^{k} =\ket{\mu_k}\bra{\mu_k},
  M_{2}^{k} = I-M_{1}^k\}$, where  $\mu_k= r e^{i 2\pi (k-1)/n}/\sqrt{n}$, on the $k$th output mode for $k=1,\dots,n$. If we detect the second outcome ($M_2$) from the $k$th measurement, we can certainly say that the input coherent state was not coherent state $\ket{r e^{i 2\pi (k-1)/n}}$. Therefore, the input coherent state can be unambiguously identified if we detect the second outcome ($M_2$) from $n-1$ measurements. The corresponding probability is 
  \begin{equation}
      P_{\rm LON}^{(n)}=\prod_{k=1}^{n-1}\Bigg(1-\exp\!\Big(-\frac{r^2}{n} |e^{i2\pi k/n}-1|^2\Big)\Bigg).
  \end{equation}
  For the case $n=2$, we can see that this method is optimal, that is, $P_{\rm LON}^{(2)}=P_{\rm D}^{(2)}$. However, this method is not optimal for $n>2$, which can be seen by comparing the approximation of $P_{\rm LON}^{(n)}$ for $r\ll 1$,
  \begin{equation}
      P_{\rm LON}^{(n)}\approx \frac{n^2 r^{2(n-1)}}{n^{n-1}},
  \end{equation}
  where we used $\prod_{k=1}^{n-1}|e^{i2\pi k/n}-1|^2=n^2$, with Eq.~\eqref{eq:AppOptProb-USD}~\cite{vanEnk2002}. 

\end{document}